# What Bell Did

Tim Maudlin

Department of Philosophy, NYU, 5 Washington Place, New York, NY 10003, USA

E-mail: twm3@nyu.edu

Abstract
On the 50th anniversary of Bell monumental 1964 paper, there is still widespread misunderstanding about exactly what Bell proved. This misunderstanding derives in turn from a failure to appreciate the earlier argument of Einstein, Podolsky and Rosen. I retrace the history and logical structure of these arguments in order to clarify the proper conclusion, namely that any world that displays violations of Bell's inequality for experiments done far from one another must be non-local. Since the world we happen to live in displays such violations, actual physics is non-local.

The experimental verification of violations of Bell's inequality for randomly set measurements at space-like separation is the most astonishing result in the history of physics. Theoretical physics has yet to come to terms with what these results mean for our fundamental account of the world. Experimentalists, from Freedman and Clauser and Aspect forward, deserve their share of the credit for producing the necessary experimental conditions and for steadily closing the experimental loopholes available to the persistent skeptic. But the great achievement was Bell's. It was he who understood the profound significance of these phenomena, the prediction of which can be derived easily even by a freshman physics student. Unfortunately, many physicists have not properly appreciated what Bell proved: they take the target of his theorem—what the theorem rules out as impossible—to be much narrower and more parochial than it is. Early on, Bell's result was often reported as ruling out *determinism*, or *hidden variables*. Nowadays, it is sometimes reported as ruling out, or at least calling in question, *realism*. But these are all mistakes. What Bell's theorem, together with the experimental results, proves to be impossible (subject to a few caveats we will attend to) is not determinism or hidden variables or realism but *locality*, in a perfectly clear sense. What Bell proved, and what theoretical physics has not yet properly absorbed, is that the physical world itself is non-local.

Bell himself faced some of these errors of interpretation, and fought to correct them. In particular, he insisted that neither determinism nor "hidden variables" were *presupposed* in the derivation of his theorem, and therefore simply renouncing determinism or "hidden variables" in physics would not change the import of the theorem in any way. This observation is of critical importance, and explains why some physicists fail to see what is so profound about his result. If the theorem only spelled out consequences for *deterministic* theories or for *"hidden variables"* theories, then the typical physicist could just shrug: since the standard understanding of quantum theory rejects both "hidden variables" and determinism, one could seemingly conclude that Bell's result says nothing about quantum theory at all. Physicists who had already abandoned "hidden variables" and determinism before reading Bell's work would seemingly find in it only further confirmation of what they had already concluded.

It is still not unusual to find people who claim that Bell's theorem forecloses the very possibility of any deterministic "completion" of quantum theory. That is, many believe that Bell accomplished what von Neumann was often supposed to have done: shown that no deterministic "hidden variables" theory could, in principle, reproduce the empirical predictions of quantum mechanics. In particular, since the most widely known deterministic "hidden variables" theory is the "pilot wave" theory of de Broglie and Bohm, it has sometimes been said that Bell's work (and the empirical verification of violations of his inequality) has *refuted* this theory.

The wrong-headedness of all this is indicated by the fact that Bell was the perhaps the strongest and most vocal advocate of the pilot wave approach in history: even stronger and more vocal than de Broglie and Bohm. From "On the impossible pilot wave":

> Why is the pilot wave picture ignored in text books? Should it not be taught, not as the only way, but as an antidote to the prevailing complacency? To show that vagueness, subjectivity, and indeterminism, are not forced on us by experimental facts, but by deliberate theoretical choice? [1] 160

And again at the end of "Six possible worlds of quantum mechanics":

> We could also consider how our possible worlds in physics measure up to professional standards. In my opinion, the pilot wave picture undoubtedly shows the best craftsmanship among the pictures we have considered. But is that a virtue in our time? [1] 195

How widespread is the misunderstanding of what Bell proved today? Here is one striking piece of evidence. On March 4, 2013, *Physics World*, the official magazine of the Institute of Physics, posted a video online titled "Why did Einstein say 'God doesn't play dice'?"[2]. The video "explains" why Einstein did not accept quantum mechanics and why Einstein is now known to have been wrong. Here is a complete transcript:

> So the everyday world is governed by classical mechanics, and that is completely deterministic. Now you might be thinking: "why do we use probabilities when we, say, throw a dice[sic]?". Well, that's actually just because of lack of knowledge about what's going on: you don't know exactly how it was thrown, where the imperfections in the table are, etc.. But in theory, if you knew all that you could work out exactly what number's going to come up every time and win all the money in the world when you are gambling.
>
> The problem is when you go to quantum mechanics—the level of the atomic and subatomic—things are a bit different, and a bit weird and sort of fuzzier let's say, because all you can actually work

> out is the probability that the thing you are looking at is going to be in a certain state. And that's not because of a lack of knowledge, that's just how it is. Now Einstein didn't like this idea at all, and that's why he said God, who is presumably all-knowing, doesn't play dice. Except quantum mechanics agrees with experiments extremely well. And even people who have tried to make so-called hidden variables theories, which means that there is sort of a secret proper state that we don't know about, can't seem to do it. Bell's theorem actually says it's impossible. So I'm afraid, Einstein, it looks like God does play dice.

There is no diplomatic way to accurately characterize the content of this video so I will stick with the plain truth: everything said here about Einstein's dissatisfaction with the standard understanding of quantum mechanics, and the bearing of Bell's work on these questions is completely wrong. What this account displays is complete incomprehension of what Bell did, what Einstein thought, what the situation with respect to deterministic "hidden variables" theories is. Indeed, the main point of Bell's theorem, and the main locus of Einstein's objections to quantum theory, namely *non-locality*, is not even mentioned.

I have no brief to single out *Physics World* here. Many physicists would take the content of the video to be uncontroversial. Indeed, the venue and presentation of this video indicates, accurately I'm afraid, that the physics community as a whole, on average, has not merely no understanding of what Bell did but the opposite of understanding: they think Bell proved something that he not only did not prove but vociferously rejected. That this situation should obtain a full half-century after the publication of Bell's result is, simply put, a scandal.

The misunderstanding of Bell and the misunderstanding of Einstein are not unrelated. Indeed, Einstein's fundamental complaint about the standard interpretation of quantum theory has been systematically distorted in just the way the video suggests, as if what Einstein could not abide was indeterminism. "Der Herrgott würfelt nicht" is probably the single most widely cited declaration of Einstein associated with his rejection of the standard account of quantum theory. But *indeterminism per se is not what bothered Einstein at all*. It is rather what indeterminism together with the predictions of quantum theory itself (in particular the EPR correlations) further imply that Einstein could not accept. Here is a direct statement by Einstein[3] 68:

> It seems hard to sneak a look at God's cards. But that he plays dice and uses "telepathic" methods (as the present quantum theory requires of him) is something that I cannot believe for a moment.

Note the second part of Einstein's concern: not merely that God plays dice but that he "uses 'telepathic' methods". This is, of course, the "*spukhafte Fernwirkung*" ("spooky action-at-a-distance") that Einstein is also known to have railed against. A careful reading of Einstein makes clear that it is the spooky action-at-a-distance, i.e. the *non-locality*, implicit in the standard account of quantum theory that bothered him, not the indeterminism *per se*. Einstein did not look for a deterministic underpinning of quantum mechanical predictions because he was wedded to determinism, he did so because he was wedded to *locality*, and he was the first to recognize that in quantum theory indeterminism can further imply non-locality.

The argument linking the two (and raising the question of the "completeness" of the quantum-mechanical description of the system, and hence "hidden variables") was, of course, the Einstein-Podolsky-Rosen (EPR) argument. It is impossible to appreciate what Bell did without first appreciating what EPR established. Indeed, one of the main reasons Bell is misunderstood today is that he both appreciated the import of the EPR argument and expected his readers to have appreciated it as well. The title of the paper we celebrate in this volume is "On the Einstein-Podolsky-Rosen paradox", and Bell explicates the point of the EPR paper in its first paragraph. He takes it that the reader already understands what the EPR argument has established: that the standard quantum mechanical account of the EPR correlations is not local, but that local physical accounts of *those particular* correlations are nonetheless possible. In fact, such local physical accounts must *also* be deterministic, so the quest for locality *implies*, in the context of discussing the EPR correlations, a quest *also* for determinism. But locality is the goal, and determinism just the necessary means.

So the misunderstanding of Bell and the misunderstanding of Einstein, both illustrated by the *Physics World* video, are not merely similar or analogous: they are conceptually linked. If one does not understand what the EPR argument had already done (which Bell only very briefly recapitulates) you cannot understand his 1964 paper. We must start with Einstein and dispel the confusions there.

Bell explicitly voiced his distress at the misunderstanding of EPR. This long passage, all of which is significant, is from "Bertlmann's socks and the nature of reality" [1] 143-4:

> It is important to note that to the limited degree that *determinism* plays a role in the EPR argument, it is not assumed but *inferred*. What is held sacred is the principle of "local causality" or "no action at a distance". Of course, mere *correlation* between distant events does not itself imply action at a distance, but only correlation between the signals reaching the two places. These signals, in the idealized example of Bohm, must be *sufficient* to determine whether the particles would go up or down. For any residual undeterminism could only spoil the perfect correlation.
>
> It is remarkably difficult to get this point across, that determinism is not a *presupposition* of the analysis. There is a widespread and erroneous conviction that for Einstein determinism was always *the* sacred principle. The quotability of his famous 'God does not play dice' has not helped in this respect. Among those who had great difficulty in seeing Einstein's position was Born. Pauli tried to help him in a letter of 1954:
>
>> …I was unable to recognize Einstein whenever you talked about him in either your letter or your manuscript. It seemed to me as if you had erected some dummy Einstein for yourself, which you then knocked down with great pomp. In particular, Einstein does not consider the concept of 'determinism' to be as fundamental as it is frequently held to be (as he told me emphatically many times)…he

> *disputes* that he uses as a criterion of admissibility of a theory "Is it rigorously deterministic?"...he was not at all annoyed with you, but only said you were a person who will not listen.

Born had particular difficulty with the Einstein-Podolsky-Rosen argument. Here is his summing up, long afterwards, when he edited the Born-Einstein correspondence:

> The root of the difference between Einstein and me was the axiom that events which happen in different places A and B are independent of one another, in the sense that an observation on the state of affairs at B cannot teach us anything about the state of affairs at A.

Misunderstanding could hardly be more complete. Einstein had no difficulty accepting that affairs in different places could be correlated. What he could not accept was that an intervention at one place could *influence*, immediately, affairs at the other.

It is impossible to read this passage and not be moved by the fact that the stubborn, almost willful, misunderstanding of Einstein has been followed by a similar almost willful misunderstanding of Bell himself, which remains to this day. How many people have asserted, over his clear and loud protests, that determinism is a *presupposition* rather than an *inference* in his own theorem? Everyone who reports Bell's result as ruling out "local, deterministic, hidden variables theories", as if the extra qualifications "deterministic" and "hidden variables" are *presuppositions of the analysis*, is doing exactly the same violence to Bell's argument that Born did to Einstein's, and indeed for the very same reason. Bell must have shaken his head ruefully at most of the physics community and thought: you are people who will not listen. At least he could have taken some comfort in having Einstein as company.

Before beginning the analysis of the EPR argument and Bell's proof, one short comment is in order. The reader may feel that my claim that there is widespread misunderstanding of both the EPR argument and the structure and content of Bell's paper should be backed up with more evidence than I have given. After all, the only concrete example I have provided is the video cited above. But even beginning to document these misunderstandings would be a Herculean task, which would eat up the meager space I have here to set the record straight. I hope that the story of how I came by the video will suffice. I did not set out to find it, and had not been apprized of it. Rather, knowing that misunderstanding of Einstein was rife and wanting an illustration, on January 9, 2014 I Googled the words "Einstein dice". This video was the fourth hit on the first page, and since it originated from *Physics World* I thought I would see what it was. The false claim about Bell's theorem at the end was a surprise bonus. That is how thick the world is with misunderstandings of Einstein and Bell. Examples are so easy to find that anyone can do it. Sadly, it is virtually certain that similar misunderstandings can be found in this very volume of papers dedicated to Bell.

Let's start with EPR.

# The Point of EPR

Discussions of the Einstein-Podolsky-Rosen argument (N.B.: *argument* not *paradox*) have always been made more difficult by the 1935 paper itself [3]. As Arthur Fine has helpfully pointed out, Einstein was not satisfied with the way the paper came out (cf. [5] 35-6). My own discussion will therefore make use of Einstein's comments in the autobiographical account he wrote in 1946, later published in *Albert Einstein: Philosopher-Scientist* [6]. I want to focus on a particularly clear and straightforward argument that can be found in the original paper leading to the conclusion that the quantum-mechanical description of a system *cannot* be a complete physical description of the system. This was, after all, the main question which the paper addressed, as evidenced by its title: "Can quantum mechanical description of reality be considered complete?". In particular, I will not consider other parts of that paper, such as the discussion of the Heisenberg uncertainty relations.

The EPR argument, of course, comes to the conclusion that the quantum-mechanical description of reality *cannot* be complete. That argument, like Bell's 1964 argument, has two premises: 1) that the predictions derivable from the quantum formalism are accurate and 2) a locality assumption. The locality assumption is not as forthrightly expressed as one might have preferred, so our main objective now is to lay it open to view and clearly articulate its contents. This locality assumption can, of course, be denied, in which case the conclusion about the incompleteness of quantum mechanics need not be accepted. But if that is the position one wishes to take, then it should be taken clearly and bluntly. One can say: "I reject Einstein's conclusion that the quantum-mechanical description is incomplete *because I reject his locality assumption*; the theory I defend is non-local in Einstein's sense". Einstein himself seemed to think that his locality assumption was so fundamental that no one would forthrightly deny it. In a letter to Born, he even suggests that the practice of physics itself cannot proceed if one denies it ([7] 170-1). That latter contention is clearly mistaken. But what is not open as a logical possibility, what is taken off the table by the EPR argument, is a position maintaining that the predictions of quantum mechanics are accurate, that the quantum-mechanical description is complete, and that the theory under consideration is local in Einstein's sense. One could, of course, try to argue that this sense of "locality" is somehow not appropriate. But a close consideration shows that "locality" is a perfectly intuitive and accurate name for the condition.

### The EPR Criterion of an Element of Reality

One reason that the role of a locality assumption in EPR has been so murky is that many readers of that paper have focused on what the paper does explicitly provide, namely the famous EPR "criterion of an element of physical reality" [4] 777-8:

> The elements of the physical reality cannot be determined by *a priori* philosophical considerations, but must be found by an appeal to results of experiments and measurements. A comprehensive definition of reality is, however, unnecessary for our purpose. We shall be satisfied with the following criterion, which we regard as reasonable. *If, without in any way disturbing a system, we can predict with certainty (i.e. with probability equal to unity) the value of a physical quantity, then there exists an element of physical reality corresponding to this physical quantity*. It seems to us that this criterion, while far from exhausting all possible ways of recognizing a physical reality, at least provides us with one such way, whenever the conditions set down in it occur. Regarded not as a necessary, but merely as a sufficient condition of reality, this criterion is in agreement with classical as well as quantum-mechanical ideas of reality.

If I could make one change to the EPR paper in retrospect it would be to alter the characterization of this criterion. The authors call it "reasonable" and "in agreement with classical as well as quantum-mechanical ideas of reality", but its status is actually much stronger than that: the criterion is, in the parlance of philosophers, *analytic*. That is, this criterion follows just from the very meanings of the words used in it. The difference is this: one *can* coherently (but not reasonably!) deny a merely reasonable claim, but one can't coherently deny an analytic proposition. Some people, hell-bent on denying the conclusion of the EPR argument, have taken to thinking they just have to reject this EPR criterion. If this can be done, then the conclusion need not be accepted. It is sometimes rhetorically suggested that the criterion itself is appropriate only in the setting of classical physics, which EPR directly deny, as the passage shows. More strongly, it has even been suggested that quantum mechanics somehow requires a rejection of classical *logic*. This illustrates only how far desperation will drive some people. There is not space to discuss this dead-end gambit here.

How is the criterion analytic? Just consider the terms used in it. What is meant, in the first case, by "disturbing a physical system"? This requires changing the physical state of a system: if the physical state is *not* changed, then the system has not been disturbed. If I do something that does not disturb the physical state of a system, then after I am done the system is in the same physical state (or lack of physical state, if that makes sense) as it was before I did whatever I did. So suppose, as the criterion demands, that I can *without in any way disturbing a system* predict with certainty the value of a physical quantity (for example, predict with certainty how the system will react in some experiment). Then, first, there must be some physical fact about the system that determines it will act that way. That is just to say that the physical behavior of a system depends on its physical state: if a system is certain to do something physical, then *something* in its physical state entails that it will do it. So determining that the system is certain to behave in some way is determining that some such physical state (element of reality) obtains. Second, if the means of determining this did not disturb the system, then the relevant element of reality obtained *even before* the determination was made, and indeed obtained

*independently of the determination being made.* Because, as we have said, the means of determination did not (by hypothesis) disturb the system.

Now suppose, as the criterion postulates, I am even *in a position* to determine how the system will behave without disturbing it. That is, even if I don't happen to make the determination, suppose that the means exist to do so (without disturbing the system). Then, by just the same argument, there must already be some element of reality pertaining to the system that determines how it will behave. For by assumption, my performing or not performing the experiment makes no difference to the system itself.

The EPR "criterion of existence of an element of reality", then, is just not the sort of thing that can coherently be denied. If one does not like the EPR conclusion, then one has to do something other than trying to deny the criterion.

Of course, there *is* something else one can do: not deny the criterion itself but rather *deny that the criterion applies in the situation EPR discuss*. After all, applying the criterion requires accepting that some physical operation does not, in fact, disturb the system in question. If an operation does disturb the system, then the criterion tells us nothing about the elements of reality. This does not "refute" the criterion: it just renders it irrelevant to the situation.

Disturbance and "Measurement"

Before turning to the EPR situation, it might profit us to reflect a moment on what has often been said about "disturbance" in relation to quantum theory. Early accounts of the Heisenberg uncertainty relations often suggested that those relations were grounded in the quantum-mechanical necessity that a "measurement" disturb the system measured. This necessity, in turn, was often somehow supposed to follow from the finite value of Planck's constant. The tacit reasoning seemed to be this: to measure a system one must interact with it, and on account of Planck's constant, the "action" of the interaction cannot be reduced below a certain point. (See, for example, Heisenberg's own account of the "Heisenberg microscope in [8]20 ff.) In contrast, it is said, in classical physics it is possible to reduce the disturbance created by a "measurement" to any degree desired (or at least to account for the disturbance, such as the change in temperature that may result from putting a thermometer into a liquid). In the limit, classical physics allows one to determine any quantity without disturbing the system, while in quantum theory this is impossible.

These claims are typically made with no justification, and indeed it is very hard to see how they could be justified. For example, the phrase "classical physics" certainly covers Newtonian mechanics and Maxwellian electrodynamics, but presumably many other possible theories as well. By what *general* argument could one establish that in any such theory any quantity can be measured to arbitrary accuracy with arbitrarily small disturbance of the system? Measurement requires a physical interaction between a system and an apparatus, and the physical analysis of that interaction obviously requires that one have details of the apparatus and details of the interaction. From what premises could a *general* proof of this proposition even be attempted?

More intriguingly, the exact opposite of the commonly-made claim above is also commonly made! Discussions of the famous Elitzur-Vaidman bomb problem commonly assert that the example shows how one can use quantum mechanics to do something that *cannot* be done in classical physics, namely determine that a bomb is not a dud without disturbing it in any way (and hence without setting it off). Indeed, Elitzur and Vaidman's paper [9] is titled "Quantum-mechanical interaction-free measurements". The question of how "measurements" can, or must, disturb systems is ripe for some careful discussion.

Locality in the EPR Argument

Given all of the talk about how measurements must disturb systems, how can EPR be so confident that they have described a situation in which their criterion properly applies? It is here that *locality* comes into the argument. EPR attempt to cut the Gordian knot: instead of finding out about a system by a procedure carried out in its vicinity, they point out that quantum mechanics, in some circumstances, allows one to acquire information about the system by a procedure carried out *arbitrarily far away.* The fundamental assumption—the assumption upon which the whole of the EPR argument depends—is that a procedure carried out arbitrarily far away from a system cannot disturb the system. Extreme spatial separation is, as it were, an insulator against disturbance. If this is correct, then procedures that can be carried out arbitrarily far from the system in question cannot disturb it, and the EPR criterion properly applies.

The whole of the EPR argument follows from 1) the criterion for an element of reality, which as we have seen is analytic (given the meaning of "disturb"), 2) the predictions of quantum mechanics itself, in particular the perfect EPR correlations (for positions and momenta in the actual case they discuss, and for spin values in the same direction in Bohm's version that Bell uses), and 3) the claim that experiments carried out on one particle do not disturb the physical state of the other. Since the first principle is analytic, and the predictive accuracy of quantum mechanics is taken for granted, the only premise in need of support is premise 3. The justification offered for 3 (implicitly) appeals to facts about spatial separation: if lab 1 is sufficiently far from lab 2, then the procedures carried out in lab 1 do not disturb the physical state of lab 2. Let's formalize the content of the claim with a definition:

> A physical theory is *EPR-local* iff according to the theory procedures carried out in one region do not immediately disturb the physical state of systems in sufficiently distant regions in any significant way.

This definition obviously needs to be tightened up to be precise: what do "immediately" and "sufficiently distant" and "in any significant way" come to? But it is best to start with this slightly sloppy formulation, not distracted by side issues. For suppose a theory *is not* EPR-local in this sense; suppose a physical theory presents an account of the physical world that does not satisfy this loose definition. Such a theory must assert that procedures carried out in one region *do* (at least sometimes) immediately and significantly disturb the physical state of a system no matter how far away that other system is. Thus, failure of EPR-locality in a theory corresponds precisely to what Einstein constantly invoked: spooky action-at-a-distance or "telepathy". If doing something here can, according to a physical theory, immediately and significantly disturb (i.e. change) the physical state of an arbitrarily distant system, then "spooky action at a distance" is a perfectly good characterization of the situation.

Why should Einstein anticipate that the correct physical theory of the actual world will be EPR-local? Well, the standard examples of *classical* physics are certainly EPR-local, and this for two different reasons. One is the *attenuation of physical influence with distance.* For example, Newtonian gravitational theory is often characterized as a theory with action-at-a-distance: if the gravitational force exerted by A on B is instantaneous and a function of the distance between A and B, then just *moving* A by any amount will immediately change the physical state of B by altering the gravitational force on B. And this is true (if the force is instantaneous) no matter how far away B happens to be. But since the gravitational force falls off as the inverse of distance squared, the *amount* to which the physical state at B is disturbed by moving A can be made arbitrarily small by transporting B far enough away. The disturbance cannot become literally zero, but it can be made *negligible,* in just the way that one says that the gravitational influence of Pluto on the tides on earth is negligible.

The second reason that the standard examples of classical physics are EPR-local does not turn on the qualifier "in any significant way" but on the qualifier "immediately". Newtonian gravity, it is often said, acts instantaneously, and hence immediately in time. And one can write down (against the background of classical space-time structure) theories containing such instantaneous influence. But Newton himself certainly did not believe the gravitational influence to be instantaneous in this sense. Newton thought that the gravitational force between A and B is produced by some *particles* that travel between A and B. The precise nature and manner of interaction of these particles was exactly the *hypothesis* that Newton declined to *fingere* in the *Principia*. But just from the fact that the gravitational interaction is mediated by particles one can infer that it is not instantaneous: the particles would take some finite period to get from A to B. This would produce a lag time between moving A and changing the gravitational force on B. Since the situations that Newton analyzes in *Principia* involve effectively static gravitational fields (the Sun is treated as *at rest*, producing gravitational forces on the planets, for example), this lag time does not come into the analysis at the level Newton carries it out. Newton would certainly have acknowledged that in a completely precise analysis, in which the motion of the Sun itself it taken into account, the lag time would make a difference to the analysis.

So suppose we accept that gravitational forces do require some transit time from A to B, no matter how small. Then by removing B sufficiently far from A we can ensure that jiggling A will have no *immediate* effect on B: the effect on B will not occur until enough time has passed for the mediating influence to get from A to B. Operations on A at time $T_1$ could not disturb the physical state of B *in any way at all* at time $T_2$ unless ($T_2 - T_1$) is enough time for the influence to get from A to B. Spatial distance here serves not directly as an attenuator of influence through the inverse-square character of the force, it rather produces a *temporal* insulator from influence via the increased transit time. This latter sort of insulation can be cleanly implemented in a Relativistic setting, if the theory implies that events at one location cannot disturb or influence any events at space-like separation. This relatively clear version is usually discussed in the context of Bell's theorem: a theory is non-local if there is some influence between space-like separated experiments. But it is notable as well that any theory that recovers the predictions of quantum mechanics must also violate the spatial attenuation condition: the violations of Bell's inequality for electrons prepared in a singlet state, for example, is predicted to remain exactly the same no matter how far apart the relevant experiments on the two particles are.

So gravity would count as EPR-local twice over if the theory postulates some time lag for gravitational effects: there is both the time lag argument and the inverse-square attenuation argument. Similarly for classical electrostatic forces (assuming a time lag). Electromagnetism (Maxwell's equations) is a bit trickier: there is a time lag built into the equations, and in most cases also attenuation with distance. In fact *every theory in the history of physics before quantum theory was EPR-local*. So the discovery that *the world* is not EPR-local (i.e. that any physical theory that makes accurate predictions cannot be EPR-local) would mark a radical break in the history of physics.

What EPR Argued

Einstein, of course, was not out to show that the physical world is not EPR-local. He rather took it for granted that the physical world is EPR-local (there is no spooky action-at-a-distance) and meant to show, from that *premise*, that the quantum-mechanical description of the world is not complete. That is, there must exist elements of reality that are not represented in the quantum-mechanical description of a system. That alone is enough to undercut any claims that quantum mechanics is, much less *must be*, a final and complete physical theory. That is the target of the EPR paper, as the title indicates.

The further conclusion that a final and complete physical theory must be deterministic *at least with respect to these particular phenomena* just comes as an additional bonus. If the world is EPR-local, and there is no spooky action-at-a-distance, then not *only* must the quantum mechanical description of a system leave out some elements of reality, but the elements that it leaves out must be sufficient, in these circumstances, to completely predetermine the outcome of the "measurement" operation. For, as Bell remarks in the passage cited above, "any residual undeterminism could only spoil the perfect correlation". This further conclusion of predetermination obviously requires that the relevant correlations be perfect, which is also what is required here to apply the EPR criterion ("*we can predict with certainty (i.e. with probability equal to unity*").

Someone may object that the condition of EPR-locality as I have rendered it above is fatally flawed on account of the admitted vagueness of some of the terms in it. "Immediately", "sufficiently distant", and "in any significant way" are not perfectly precise terms, and maybe "procedure" isn't either. But precise definitions are not needed if one is only interested in a particular, concrete case. So, for example, although "procedure" is a somewhat vague term, there is no doubt that the sorts of things relevant here (i.e. the sorts of things that go on when one performs a "spin measurement") count as a "procedure". The definition of EPR-locality given above is similarly only problematic if there is doubt about how to apply the terms *in the concrete situation being analyzed*. Einstein, Podolsky and Rosen were presumably sensitive to this sort of issue: that is why they were content to provide just a *criterion* (sufficient condition) for an "element of reality" rather than a *definition* (which would at least have to be necessary and sufficient). They are rather insistent about this, with good reason.

Einstein's intuitive notion of "telepathy" or "spooky action-at-a-distance" (which is just what an EPR-local theory rules out) is also described in his Autobiographical Notes, the first sentence of which Bell cites in his paper [6] 85:

> But on one supposition we should, in my opinion, absolutely hold fast: the real, factual situation of the system $S_2$ is independent of what is done with the system $S_1$ which is spatially separated from the former. According to the type of measurement which I make of $S_1$, I get, however, a very different $\psi_2$ for the second partial system…Now, however, the real situation of $S_2$ must be independent of what happens to $S_1$. For the same real situation of $S_2$ it is possible therefore to find, according to one's choice, different types of $\psi$-function. (One can escape from this conclusion only by assuming that the measurement of $S_1$ ((telepathically)) changes the real situation of $S_2$ or by denying independent real situations as such to things that are spatially separated from each other. Both alternatives appear to me entirely unacceptable.)

Since Bell was relying on the EPR argument, and we know Bell had been reading exactly this passage, it is worthwhile to spend a moment on it. First, Einstein asserts that there is a "supposition" required for his argument to go through, exactly the supposition of EPR-locality mentioned above. Further, it is clear that in this context "what is done" with $S_1$ is that the sort of experiment that is commonly called a "measurement" is carried out on it and the result of the experiment is noted. It might, as Einstein says, be any of a number of possible different "measurements": it might (as in the original EPR paper) be a "position measurement" or be a "momentum measurement", or it might, as in the Bohm singlet example, be a "measurement" of spin in any direction at all. One thing that no one disputes is that by means of carrying out any of these procedures on $S_1$, one can come to be in a position to "predict with certainty (i.e. with probability equal to unity)" how system $S_2$ will respond if the same "measurement" is carried out on it. So on the basis of this sort of procedure carried out on $S_1$ one can come to know, with certainty, a particular physical fact about $S_2$: namely how it is disposed to react to such a measurement. Now the key question for Einstein is this: did the procedure carried out on $S_1$ *influence or change or disturb or bring into existence* any part of the "real factual situation" of $S_2$? If it *did*, then that is what Einstein calls "telepathy". If it did, then Einstein concedes that the rest of his argument fails. If it did, then even in the face of the perfect EPR correlations, the quantum-mechanical description of $S_1$ and $S_2$ (and also, if one wants to make the distinction, the joint system $S_1 + S_2$) could be complete. But if the procedure carried out on $S_1$ did *not* disturb or alter or change or bring into existence the real factual situation of $S_2$, then $S_2$ must have this physical disposition to react to the "measurement" *all along*. Since the initial quantum state ascribed to $S_2$ does not ascribe it such a disposition, the initial (quantum mechanical) state must have been *incomplete*.

In this passage Einstein offers *two* possible ways to reject the conclusion of his argument: accept telepathy or reject the claim that systems spatially separated from one another even have "independent real situations". Unfortunately, Einstein never discusses this second option in detail. My guess is that Einstein couldn't even really imagine what this second option could be like. But more than that, it is not even clear to me exactly how this move is supposed to interfere with the argument. If we use the EPR criterion of an element of reality, then we get the existence of an element of reality in $S_2$ that was not represented in the quantum-mechanical description of $S_2$, so the quantum-mechanical description is not complete. So merely assuming EPR-locality gets us the incompleteness of the quantum description without any further supposition. The only thing I can think of is an opponent who denies that $S_1$ or $S_2$ have individually *any physical state at all*, so that one cannot even say that as a matter of physical fact a certain experiment was carried out on $S_1$ that had a certain outcome, all of which would have to be reflected in something about the physical state of the region where $S_1$ is located. That is, Einstein might have had some idea of a theory that, in Bell's later terminology, *postulates no local beables at all*. In such a theory, one could presumably not make any physical claims about what was done to $S_1$ and how it reacted: there just would be no physics of $S_1$ per se. It is extremely obscure how any such theory could possibly make physical sense of the laboratory operations that we *think* were carried out on $S_1$, and the sorts of reactions we *think* $S_1$ displayed to those operations. So it is very obscure how such a theory could make contact with, or explain, anything that we take to be empirical data. But since Einstein tells us no more about this option, we can only speculate.

The main point—the point essential for understanding Bell's 1964 paper (which is, after all, our target)—is that it is absolutely clear that nowhere does Einstein assume *determinism* in his argument. "No telepathy" certainly does not, on its own, entail determinism. And "spatially separated system have their own physical states" certainly does not entail determinism. But these principles *together with the EPR correlations* entail not only that quantum mechanics is not complete, but also that any complete theory (indeed, any theory just *complete enough* to represent the elements of reality at play in this experiment, assuming EPR-locality) must *also* be a deterministic theory with respect to those particular experiments.

That argument is one line: the very "element of reality" that the EPR argument proves to exist—given EPR-locality—is an element of reality defined just as whatever physical characteristic of the system it is that ensures how it would react to the measurement in question. So any system that *has* that element of reality has a physical characteristic that determines how it would react to the measurement. But that just is determinism with respect to that particular "measurement operation". And the EPR argument can be repeated for any "measurements" for which quantum theory predicts perfect correlations between the outcomes and that can be made arbitrarily far apart in space. Hence, in an EPR-local theory both the reactions to a "position measurement" and to a "momentum measurement" must be predetermined by some element of reality in the system, and in the Bohm spin example the reactions to every possible "spin measurement" must be predetermined. That is enough to get Bell's 1964 argument off the ground. Not by assuming determinism, but by assuming EPR-locality and deriving determinism.[1] Just as Bell said.

**Between EPR and Bell**

Bohr

The only way to avoid the EPR conclusion that the quantum-mechanical description of the system is incomplete, and the additional conclusion that in any physics with a complete description the results of these "measurements" will be predetermined, then, is to forthrightly deny that the physical theory is EPR-local, and to forthrightly admit that it does postulate "telepathy" in Einstein's sense. It is notable that Bohr's own response to the EPR argument sort of does this (and sort of doesn't).

Bohr begins with some boilerplate comments about quantum theory of the "uncontrollable and unpredictable finite disturbance" ilk [10] 696-7:

> Such an argumentation, however, would hardly seem suited to affect the soundness of quantum-mechanical description, which is based on a coherent mathematical formalism covering automatically any procedure of measurement like that indicated. The apparent contradiction in fact discloses only an essential inadequacy of the customary viewpoint of natural philosophy for a rational account of physical phenomena of the type with which we are concerned in quantum mechanics. Indeed, the *finite interaction between object and measuring agencies* conditioned by the very existence of the quantum of action entails—because of the impossibility of controlling the reaction of the object on the measuring instrument if these are to serve their purpose—the necessity of a final renunciation of the classical ideal of causality and a radical revision of our attitude towards the problem of physical reality. In fact, as we shall see, a criterion of reality like that proposed by the named authors contains—however cautious its formulation may appear—an essential ambiguity when it is applied to the actual problems with which we are here concerned.

---

[1] *Deriving* determinism is this way for a local theory simultaneously *derives* the condition sometimes called "counterfactual definiteness". In a deterministic theory, the physical state of a system not only determines how it will react to any procedure carried out on it, but also how it would have reacted had a different procedure been carried out.

A few preliminary remarks are apposite.

First, touting the fact that quantum mechanics provides a "coherent mathematical formalism automatically covering any procedure of measurement like that indicated" is rather ironic since the EPR argument uses exactly that formalism and completely trusts in its accuracy: it is that formalism which predicts the perfect EPR correlations that the argument presupposes. More telling is the meaty sentence about the uncontrollable and unpredictable finite interaction between the object and the measuring apparatus, which shows that Bohr has failed to understand the EPR situation at all. The beauty of the EPR set-up is that it doesn't matter at all how extensive and violent the interaction *in the vicinity of $S_1$* is, or how that might make impossible further predictions about *how the particle in $S_1$ will behave in later measurements on it*. It is exactly because the interaction with $S_1$ is being used to *provide information about $S_2$* that all of the concerns about "finite interaction" and "controlling the reaction" are completely irrelevant *if the theory is EPR-local*. In an EPR-local theory you can blow up $S_1$ with a thermonuclear device if you like (that counts as a "finite interaction"!), still *that can make no immediate physical difference to the state of $S_2$*. That is what underlies the rather commonsense idea that you don't have to worry about the immediate effects of thermonuclear explosions if they are far enough away. Bohr has apparently failed to appreciate the way that spatial separation is being used as an insulator against disturbance (via EPR-locality) in the argument. Or, more accurately, he is just recycling talking points written for completely non-EPR situations (does "measuring the position" of a single particle disturb the momentum *of that very particle*?) in a context where they have become irrelevant.

When Bohr comes back to the "reality criterion", the discussion gets very interesting. He does not actually deny the criterion, which, to recap, I have argued is analytic and cannot be coherently denied. Rather, Bohr takes the other route out: he accepts the criterion but argues that it does not apply in this case because there is, after all, a disturbance of the second system due to the procedures carried out on the first. That is, Bohr's words suggest that he is *granting* that quantum mechanics is in fact not EPR-local. But having said it, he then tries to also take it back [10] 700:

> From our point of view we now see that the wording of the above-mentioned criterion of physical reality proposed by Einstein, Podolsky and Rosen contains an ambiguity as regards the meaning of the expression "without in any way disturbing a system." Of course there is in a case like that just considered no question of a mechanical disturbance of the system under investigation during the last critical stage of the measuring procedure. But even at this stage there is essentially the question of an influence on the very conditions which define the possible types of predictions regarding the future behavior of the system. Since these conditions constitute an inherent element of the description of any phenomenon to which the term "physical reality" can be properly attached, we see that the argumentation of the mentioned authors does not justify their conclusion that quantum-mechanical description is essentially incomplete. On the contrary this description, as appears from the preceding discussion, may be

> characterized as a rational utilization of all possibilities of unambiguous interpretation of measurements, compatible with the finite and uncontrollable interaction between the objects and the measuring instruments in the field of quantum theory.

So there *is* a disturbance and there isn't. What EPR have in mind in their criterion is exactly a "mechanical" disturbance in the sense that *the physical state of $S_2$ changes*, but not necessarily "mechanical" in the sense of the "mechanical philosophy", i.e. by contact action. That's exactly why Einstein calls it "telepathy" or "action-at-a-distance": it is a change in physical state that is not produced by contiguous actions. If Bohr is denying that the change is "mechanical" in the first sense, then the EPR criterion still applies, and the theory is incomplete. If he is merely denying it in the second sense then he is just agreeing with Einstein: it is telepathy.

Of course, Bohr wants to say there is a change, an essential change, in $S_2$ on account of the procedure carried out on $S_1$, just not a change in its physical condition, and not (apparently) due to any "finite interaction" between the procedure carried out on $S_1$ and $S_2$. So what kind of change is it? The rest of the words are simply incomprehensible. How could the procedure carried out on $S_1$ have any relevance at all for defining "the possible types of predictions regarding the future behavior of" $S_2$? The possible *types* of predictions about the behavior of $S_2$ are predictions like this: "I predict that if you 'measure the x-spin' (i.e. carry out such-and such a procedure) of this particle you will get an up result" or "I predict that if you 'measure the y-spin', you will get a down result". The *language* in which these predictions are couched is perfectly clear, irrespective of what is done with $S_1$. Of course, finding out what happened to $S_1$ can provide information about which of these predictions is likely, or even certain, to be correct! But that is EPR's whole point, not something they overlooked.

Schrödinger

Schrödinger, in contrast, took the first steps leading from EPR to Bell's 1964 result. Part of Schrödinger's famous "cat" paper [11] contains his reflections on EPR. Indeed, Schrödinger indicates that the appearance of the EPR paper motivated him to write the "cat" paper, which he is uncertain even how to characterize: as a "report" (*Referat*) or a "general confession" (*Generalbeichte*). Schrödinger is acutely aware of the bind that EPR has created. He describes two systems in a maximally entangled state, and notes that the outcome of *any* measurement made on one can be accurately foretold by an appropriate experiment carried out on the other, even though the two systems may have been separated from each other. He amusingly analogizes the situation to having a collection of students, each of whom may be asked any of a set of questions, whose answers can be found in a book. The choice of question is at the free disposition of the schoolmaster, and whenever a question is asked and the book is then consulted, the answer turns out to be correct. But, for a given student (and for the book as well!), it is a one-shot deal: having answered any one question, the student is no longer able to answer any further question correctly.

If any student so prepared is able to answer the first question (which may be randomly chosen) correctly, it follows (writes Schrödinger) that he initially *knows*

the answers to *all* the questions that could be asked [11] 164:

> But let us once more make the matter very clear. Let us focus attention on the system labeled with the small letters p, q and call it for brevity the "small" one. Then things stand as follows. I can direct *one* of two questions at the small system, either that about q or that about p. Before doing so I can, if I choose, procure the answer to *one* of these questions by a measurement on the fully separated other system (which we may regard as auxiliary apparatus), or I may intend to take care of this afterward. My small system, like a schoolboy under examination, *cannot possibly know* whether I have done this or for which questions, or whether and for which I intend to do it later. From arbitrarily many pretrials I know that the pupil will correctly answer the first question that I put to him. From that it follows that in every case he *knows* the answer to both questions. That the answering of the first question so tires or confuses the pupil that his further answers are worthless changes nothing at all of this conclusion.

We again get the EPR. If the particle "knows" (i.e. is physically disposed to give) the correct answer to either question, then there must be an "element of reality" associated with the particle that carries the information about what that answer is. That is, this sure-fire disposition of the particle must derive from some physical characteristic of it. To deny this is not to deny *realism* or a *realist view*, as some have later seemed to claim, but to deny *physics* altogether. The disposition is there, Schrödinger insists. That disposition must be grounded in the physical state of the particle. What else can it be grounded in?

That Schrödinger appreciated the dilemma posed by EPR is clear at the end of this section. Having noted the perfect correlations between the two systems for both p and q (the measured value of p is always the negative of P and the measured value of q is always the same as Q in this calibration), Schrödinger first infers definite dispositions for each system with respect to each property (as we have seen) and then takes a new step. By measuring the position on the "small" system and the momentum on the "big" system one could get the results, e.g., q = 4 and P = 7. Then using the perfect EPR correlations in both directions, we can determine the dispositions of each system with respect to *both* possible measurements: q = 4 *and* p = –7, and also Q = 4 *and* P = 7 [11] 164.

> There is no doubt about it. Every measurement is for its system the first. Measurement on separated systems cannot directly influence each other—that would be magic. Neither can it be by chance, if from a thousand experiments it is established that virginal measurements agree.
>
> The prediction catalog q = 4, p = –7 would of course be hypermaximal.

So where Einstein presents the EPR dilemma as telepathy or incompleteness, Schrödinger presents the same options as magic or hypermaximality. But that is just different terminology for the same problem. By "hypermaximal" Schrödinger means that there are more definite quantities specified in the state than are allowed by the uncertainty relations. Indeed, much of Schrödinger's presentation of quantum theory turns on counting independent degrees of freedom in a system, the general postulate being that a quantum system has half the number of degrees of freedom as the corresponding classical system. So while in a classical system both the position and velocity of a particle always have definite values, in a quantum system at most one does. And in an entangled system, definite values for properties of the component parts can be traded off for definite conditionals such as "The spin of particle 1 will come out opposite to that of particle 2, no matter what direction they are measured in". If a quantum mechanical system is in a state that makes these conditionals true, then neither particle can have a definite spin in any direction. Adding a definite spin in any direction would by hypermaximal.

Schrödinger does not explicitly say that he rejects magic in favor of "hypermaximality", but it seems implicit in how he proceeds that he does so. In fact, the next section (Section 13) makes a new observation: not only do the perfect correlations imply (absent magic) that each particle has a definite disposition with respect to both position measurements and momentum measurements, the EPR state is so highly "entangled" that the same argument holds for *every possible measurement that can be carried out on either particle individually*. The complete "prediction catalog" must specify how every possible single-particle observation would come out: infinitely many such "elements of reality".

But more than that. What Schrödinger also realizes is that these various "elements of reality" cannot possibly be related to one another in the way that the mathematical relations between their associated Hermitian operators would naturally suggest. Schrödinger considers several different "questions" (i.e. "measurements"), namely those associated with the operators p, q, and $p^2 + a^2q^2$, for any positive constant a. Each of these corresponds to a single Hermitian operator, an "observable". Now suppose (as he has shown, absent magic) the system holds ready the answer to each of these questions, i.e. what the numerical result of the measurement would be. In the case of p and q, call these predetermined values p' and q'. And recall that the observed outcome of the $p^2 + a^2q^2$ "measurement" must be an odd multiple of $a\hbar$ since these are the eigenvalues of the operator. If the predetermined values of the "measurements" had to bear the same algebraic relations to one another as their associated operators do, then we would have, as Schrödinger points out, that $(p'^2 + a^2q'^2)/ a\hbar$ = an odd integer *for any positive real number a* [11] 164. But that is plainly *mathematically* impossible. So the predetermined values *cannot* bear the same algebraic relations as their associated operators.

This, of course, is the result that was put forward by von Neumann as an impossibility proof for any theory that posits "hidden variables" and recovers the quantum-mechanical predictions. Von Neumann could not have been more emphatic about the universal character of his conclusion [12] 325:

> It is therefore not, as it is often assumed, a question of re-interpretation of quantum mechanics, — the present system of quantum mechanics would have to be objectively false, in order that another description of the elementary processes than the statistical one may be possible.

But von Neumann has reached his conclusion only by raising the requirement of corresponding algebraic structure to the level of an axiom. Schrödinger, on the other hand, simply notes that the algebraic relations among the predetermined values could not be the same as those among their associated operators: nothing *impossible* about that.

Von Neumann's "proof" provides another interesting diagnostic test of how carefully and closely physicists were thinking things through. Apparently, an entire generation of physicists had been convinced by von Neumann's proof—or, more likely, just by hearing that von Neumann had somehow provided a "mathematical" proof—that no deterministic "completion" of quantum mechanics is possible. Schrödinger, aware of exactly the same mathematical result, simply accepts it as a price a "hidden variables" theory must pay. Bell, when he finally managed to get his hands on an English translation of von Neumann, found the "axiom" not merely rationally deniable but completely uncompelling, as he later said in an interview in *Omni* magazine [13] 88:

> Then in 1932 [mathematician] John von Neumann gave a "rigorous" mathematical proof stating that you couldn't find a non-statistical theory that would give the same predictions as quantum mechanics. That von Neumann proof in itself is one that must someday be the subject of a Ph.D. thesis for a history student. Its reception was quite remarkable. The literature is full of respectable references to "the brilliant proof of von Neumann;" but I do not believe it could have been read at that time by more than two or three people.
> 
> ***Omni***: Why is that?
> 
> **Bell**: The physicists didn't want to be bothered with the idea that maybe quantum theory is only provisional. A horn of plenty had been spilled before them, and every physicist could find something to apply quantum mechanics to. They were pleased to think that this great mathematician had shown it was so. Yet the Von Neumann proof, if you actually come to grips with it, falls apart in your hands! There is *nothing* to it. It's not just flawed, it's *silly*. If you look at the assumptions it made, it does not hold up for a moment. It's the work of a mathematician, and he makes assumptions that have a mathematical symmetry to them. When you translate them into terms of physical disposition, they're nonsense. You may quote me on that: the proof of von Neumann is not merely false but *foolish*.

One might suspect that Bell is operating here with the benefit of hindsight. In particular, as we are about to see, once Bohm's pilot wave paper came out it was obvious that von Neumann *had* to be wrong: Bohm did just what everyone (including von Neumann!) thought von Neumann had proven impossible. So between Schrödinger in 1935 and Bell (well before 1988) did anyone else figure out how lame von Neumann's proof was? David Wick reports the following [14] 286:

> The reader may find it strange—I certainly do—that neither Bohr nor Einstein brought up von Neumann's "impossibility proof" in their debates. But reflecting on von Neumann's membership in another generation, on his profession as a mathematician, and on the date his book appeared (only a few years prior to EPR) makes this lacuna appear less mysterious. We do know that Bohr later held many discussions with von Neumann at Princeton, and we can surmise he was not troubled by von Neumann's conclusions. We also know that, by around 1938, Einstein knew about von Neumann's theorem. Peter Bergmann told Abner Shimony that he, Valentin Bargmann (both assistants of Einstein) and Einstein once discussed it in Einstein's office. On that occasion Einstein took down von Neumann's book, pointed to the additivity assumption, and asked *why should we believe in that*.

We should pause to reflect here, as these facts shed light on the situation highlighted at the start of this paper. Apparently, as the *Physics World* video testifies, even a contemporary physicist who is presented in a mainstream venue as qualified to answer a fairly straightforward question can be completely misinformed or confused about what the relevant facts are. This might seem incredible. It might seem equally incredible that entire generations of physicists could be convinced that an important physical question had been settled definitively by a rigorous mathematical proof when in fact the proof was "silly" and "foolish" because it relied on an axiom that we have no reason to believe. But like it or not, that is the situation. Bohm's 1952 paper, of course, made the state of affairs clear: if a deterministic theory that recovers all of the relevant quantum-mechanical predictions *actually has been constructed*, then there can't be a good proof that this cannot be done! But the point is that some of the most prominent physicists of the age—Einstein and Schrödinger—knew the unimportance of von Neumann's result, and still the majority of the physics community nonetheless took it as profound and decisive.

The inverse of this has happened (and still is happening) with Bell. A great many physicists, and indeed a great many *great* physicists, simply do not appreciate what Bell proved. But while von Neumann proved nothing of significance and was credited with genius, Bell proved something of epochal importance that is often presented as merely putting an already moribund theoretical project out of its misery.

And you, dear reader, might well ask: but how can I tell? If the majority of physicists, and even great physicists, can be mistaken in their pronouncements on such a topic, who am I to trust?

The only possible answer is: trust only yourself. Read the arguments for yourself *and make sure you understand them*. It is this last point that is hard. It is easy to understand what someone *claims* to have proven: see the example from von

Neumann above. But you have to check that it really has been proven. And checking, in this case, does not mean checking the mathematics. That might be hard or might be easy, but it is straightforward. A well-known and widely accepted paper is not likely to contain a significant mathematical error. Von Neumann certainly did not make one. But mathematics alone is not physics, and mathematics alone has no direct physical consequences. Those only come through the interpretation of what the mathematical result signifies for the physics, and that requires understanding how the mathematics is being used to make physical claims. If you can't really follow that, then you should retain a healthy skepticism about the reported result, even if it happens to be widely accepted.

Bohm

We have only one last event to discuss before turning to Bell's paper. In 1952, Bohm published his version of the "pilot wave" theory [15], and we know Bell read the paper. He verified that the claims made in Bohm's work were true, so the standard story about what von Neumann proved must be false. But Bohm's work had more significance than that.

One important thing Bohm did in his earlier textbook ([16] 614) was just to present the EPR argument in terms of spin measurements on a singlet state rather than position and momentum measurements on an EPR state[2]. Schrödinger had already noted that for the EPR state the perfect correlations, which allow one to make predictions with certainty for $S_2$ by experiments performed only in the vicinity of $S_1$, exist not just for position and momentum but for the whole infinitude of

---

[2] I cannot forego pointing out that in this book, Bohm ends his discussion of EPR with a section entitled "Proof That Quantum Theory is Inconsistent with Hidden Variables", in which he states: "We conclude then that no theory of mechanically determined hidden variables can lead to *all* the results of the quantum theory. Such a mathematical theory might conceivably be so ingeniously framed that it would agree with quantum theory for a wide range of predicted results." To this he appends a footnote: "We do not wish to imply here that anyone has ever produced a concrete and successful example of such a theory, but only state that such a theory is, as far as we know, conceivable" ([16] 623). Murray Gell-Mann explains the *volte-face* between 1951 and 1952 [17] 170: "When I met David a day or two later [after promising to arrange an interview with Einstein] and started to tell him I was working on an appointment with Einstein, he interrupted me excitedly to report that it was unnecessary. His book had appeared and Einstein had already read it and telephoned him to say that David's was the best presentation he had ever seen of the case against him, and they should meet to discuss it. Naturally, when next I saw David I was dying to know how their conversation had gone, and I asked him about it. He looked rather sheepish and said, 'He talked me out of it. I'm back to where I was before I wrote the book.'"

So it was Einstein who motivated Bohm to write his 1952 paper, and that paper inspired Bell's work, leading to his result.

possible "measurable quantities" of either particle. But most of these "observables", such as the "observable" represented by the Hermitian operator that is the *sum* of the position and momentum operators, have no familiar physical significance. Indeed, it is not clear, as an experimental matter, how to go about "measuring" this observable: certainly not by measuring position and then momentum and adding the results! For the spin case, in contrast, each of these infinitely many possible observable quantities is just as easy to experimentally access as any other: it is just a matter of reorienting the analyzing magnet in space. It then becomes more natural to start thinking about what would happen when *different*, not-perfectly-correlated "observables" are measured on the two sides.

Schrödinger, of course, had already considered this possibility, but only for position on one side and momentum on the other. And in this case, the results on the two sides are statistically independent: there is no correlation at all. The result on $S_1$ provides no information about what the result on $S_2$ will be. To get to Bell's result in this setting, one needs to consider the intermediate cases, where the outcomes on the two sides are neither perfectly correlated nor perfectly uncorrelated. Bohm's translation of this problem to spin made this easier.

But even Bohm's paper of 1952, with its concrete refutation of von Neumann (and of Bohm 1951: see the last footnote) did not make Bell's discovery easy. If it had, presumably Bohm would have done it in 1952. The tinder was set, but still required the spark of genius to set it aflame. That took over a decade. Perhaps without Bell, we would still be waiting today.

**What Bell Did**

Having set the stage, our most important section can be short. Bohm's "pilot wave" theory assured Bell that von Neumann's, and all the other so-called "no hidden variables" proofs, had to contain errors. They claimed that no deterministic theory could replicate the predictions of quantum theory, but Bohm's theory gave them the lie. So there was a purely diagnostic question: where did these proofs go wrong, and what sorts of deterministic theories could reproduce the quantum-mechanical predictions?

The first part of his investigation into this question was the paper "On the problem of hidden-variables in quantum mechanics" [18]. This paper appeared later than "On the Einstein-Podolsky-Rosen Paradox", but was written earlier. Bell had carefully analyzed the pilot wave theory to see how it ticks, and found its dynamics to have a "grossly non-local character" [1] 11. More specifically:

> So in this theory an explicit causal mechanism exists whereby the
> disposition of one piece of apparatus affects the results obtained with
> a distant piece. In fact the Einstein-Podolsky-Rosen paradox is
> resolved in the way which Einstein would have liked least.

This comment demonstrates that Bell took the recovery of locality to be of higher importance to Einstein than the recovery of determinism, just as Einstein himself insisted.

Bell's comment is worthy of careful consideration. Einstein, he knew, hated the "telepathy" or "spooky action-at-a-distance" that he repeatedly and vociferously attributed *to standard quantum mechanics*. In *that* theory, Einstein said, God both plays dice *and* "uses telepathic methods". And it is exactly in the EPR setting, with perfect correlations between the outcomes of distant systems, that this is so obvious as to slap you in the face. If the outcome of one experiment is *not* predetermined, if it is a matter of pure and irreducible *chance*, then how can the *distant* system "know" how that random choice was made except by telepathy? The distant system always behaves appropriately, like Schrödinger's schoolboy giving the right answer. Indeed, the behavior of the distant system *provides information about how the local system has behaved or will behave*. Given just the initial set-up of an EPR-Bohm experiment, for example, we have no idea how a distant x-spin measurement will come out: either outcome is equally likely. But observations of our local particle (measuring *its* x-spin) informs us with perfect accuracy how the distant measurement came out. If there was really any chanciness involved, *how in the world did the local particle get the information that we can extract from it*? Telepathy!

    This, it seems to me, is the nub of all the present disputes and misunderstandings about the significance of Bell's theorem. On one telling, Bell's theorem spelled the end of the "hidden variables" approach, because it showed that the *price* for postulating hidden variables had to be accepting non-locality. And no right-minded person would do *that*: better stick with a *local* but *indeterministic* theory. But that gets the entire dialectical situation upside-down and backwards. What upset Einstein about the *standard* theory was that it is obviously both indeterministic *and* non-local. The EPR experiment allowed him to show that it *must* be non-local if it is indeterministic. And, as Bell saw, it was the non-locality of the theory that Einstein hated more than its indeterminism. Einstein would be willing to accept indeterminism, but not non-locality.

    The non-locality of any *indeterminstic* theory that has to recover EPR-like perfect correlations is later illustrated by Bell with the delightful example of Bertlmann's socks. As he says, there is nothing at all *prima facie* puzzling or surprising about the correlation: Bertlmann has on different colored socks at any time during the day because he put on different colored socks when he got up. And seeing one sock provides information about the other: no big deal.

    But what if, rather perversely, one were to try to insist that when he put the socks on in the morning neither sock had any particular color? When asked why you always *see* a color when you *look at* a sock (make a "color measurement") you are told that the very act of looking *brings the color into existence*, and does so in a *completely indeterministic way*: given the initial state of the sock, any one of a number of different colors might show up. Well, all this is very weird and bizarre, but not in any way *non-local*. But now, suppose we note that the observed socks are also always different colors, no matter how far apart the observations are made. Now we have telepathy: observing sock A not only somehow brings its color into existence, it *also* somehow has the effect of changing the *other* sock so that, when observed, it *cannot* assume that same color. The trivial locality of the normal deterministic account has been converted into a weird telepathy in the indeterministic account. Also note that if one is unable to *control* which color one of these weird telepathic socks will suddenly display then they cannot be used to send signals, so the telepathy involved has nothing to do with signaling.

So for Einstein, deterministic theories were not primarily desirable because they are deterministic: they were desirable because they held out the only hope of exorcising the spooks from standard quantum theory. And Bohm's theory, deterministic as it was, dissatisfied Einstein because the determinism there did no such thing. The question left for Bell is whether *some other* theory could get rid of the non-locality. We know from EPR that only a deterministic (and hence "hidden variables") theory could possibly work (or at least, a theory deterministic in EPR-like situations), so the search had already been narrowed to that sector of possible physical theories. Einstein had figured that out in 1935, and Bell saw it equally clearly. What "On the Einstein-Podolsky-Rosen Paradox" did was not extinguish all hope for hidden variables or for determinism (again, Bell was a strong *advocate* of the "pilot wave" approach) but extinguish all hope for locality.

EPR had already ruled out local indeterministic theories, and "On the Einstein-Podolosky-Rosen paradox" ruled out local deterministic theories. Between the two of them, they licked the platter clean. Locality must be abandoned.

There is no point in my reproducing Bell's argument in detail in this venue. Once we have established, via the EPR argument, that in any local theory each particle in a singlet state must have definite spin values in all directions, we only need to consider some very large collection of singlet pairs. In any such large collection there will be precise values for the correlations between the spins on the two sides in any pair of directions. It is a plain mathematical fact that no such collection can contain correlations that match the quantum-mechanically predicted correlations for *outcomes of spin measurements* in those directions. In the spin case, the spins in the same directions of both side must be perfectly (anti-)correlated, and the spins for any pair of orthogonal directions must be completely uncorrelated, and, in general, the spins for an angular difference θ between the directions have correlations proportional to $\cos^2(\theta)$. The only additional assumption of Bell's proof is that (in a local theory) the *observed* correlations are (nearly) identical to the correlations in the whole collective, i.e. that the observed spin values are a random sample from the total collection. The rest is just the Law of Large Numbers. Just like von Neumann, Bell made no mathematical mistake. But unlike von Neumann, he also made no *conceptual* mistake. He does show that any theory that reproduces the predictions of standard quantum theory cannot be local, and *a fortiori* "standard quantum theory" (which surely reproduces those predictions!) cannot be local.

One notable advance since Bell's original work is the elimination of the appeal to the Law of Large Numbers. The beautiful example of Greenberger, Horne and Zeilinger [19] shows that for certain entangled triple of particles, no local theory can guarantee quantum-mechanically acceptable outcomes of measurements for all possible measurement situations in *any single run* of the experiment. In this example, the statistical considerations drop out altogether.

The resistors against Bell have been at least as numerous as the embracers of von Neumann once were. And part of that resistance has been to propose that in some *important* or *proper* sense of "local" standard quantum theory *is* local, and all Bell has shown is that deterministic (or "hidden variables") theories cannot be local in this important sense, while standard quantum theory can be. One prominent suggestion for this proper and important sense of "local" has to do with *signaling*. In this sense, a physical theory is non-local just in case one can specify how to use the physics to send useful, interpretable signals faster than light.

But this line of argument simply makes no contact at all with what the topic of discussion has been! That is certainly not the sense of "local" that *Einstein* had in mind. Einstein was perfectly well aware that one cannot use the EPR correlations to send signals of any sort from one side to the other, but he insisted that it showed the manifest non-locality of standard quantum mechanics nonetheless. For, as he argued, standard quantum mechanics must hold that the operation on one side *does* disturb the real physical state on the other, in order for the reality criterion not to apply. And the reality criterion had better not apply if one is to maintain that the quantum-mechanical description is complete.

It is clear that this was Bell's understanding of the situation as well. "On the Einstein-Podolosky-Rosen paradox" begins thus [1] 14:

> The paradox of Einstein, Podolsky and Rosen was advanced as an argument that quantum mechanics could not be a complete theory but should be supplemented by additional variables. These additional variables were to restore to the theory causality and locality. In this note, that idea will be formulated and shown to be incompatible with the statistical predictions of quantum mechanics. It is the requirement of locality, or more precisely, that the result of an experiment on one system be unaffected by operations on a distant system with which it has interacted in the past, that creates the essential difficulty.

Note that Einstein's aim was to *restore* causality (i.e. determinism) *and locality*, features both *absent* in the standard theory. And note also that the real sticking point is locality. Determinism can, in fact, be restored while keeping the same predictions, as the pilot wave theory shows. But locality—in the sense Bell articulates, which is Einstein's sense—cannot be restored *even* in a deterministic theory. Locality must be abandoned in either case.

Other resistors of Bell have tried other gambits, too numerous to recount here. A thorough and very careful analysis of many misunderstandings of Bell can be found in Goldstein et al. [20].

This, as I initially claimed, was the great achievement of Bell. He taught us something about the world we live in, a lesson that will survive even the complete abandonment of quantum theory. For what cannot be reconciled with locality is an observable phenomenon: the violations of Bell's inequality for "measurements" performed at arbitrary distances apart, or at least at space-like separation. And this phenomenon has been verified, and continues to be verified, in the lab. Neither indeterministic nor deterministic theories can recover these predictions in a local way. Non-locality is here to stay.

**Caveats**

There are still a few small technical points and one large theoretical one.

There are niggling questions that can arise from the noisiness of actual experiments. The clear statistical predictions derived from theory are never precisely replicated in the lab. This creates some experimental loopholes, which depend on the predictions of quantum theory being *wrong*, and would require that apparent violations of the inequality evaporate as experimental technique gets better. This was always a thin and implausible reed, which arguably no longer even exists.

There is the so-called idea of "superdeterminism". Recall Schrödinger's class of identically prepared students. We are told they can *all* answer *any* of a set of questions correctly, but each can only answer one, and then forgets the answers to the rest. It's an odd idea, but we can still test it: we ask the questions at random, and find that we always get the right answer. Of course it is *possible* that each student only knows the answer to *one* question, which always happens to be the very one we ask! But that would require a massive coincidence, on a scale that would undercut the whole scientific method. Or else we are being *manipulated*: somehow we are led to ask a given question only of the rare student who knows the answer. So we switch our method of choice, handing it over to a random number generator, or the throw of dice, or to be determined by the amount of rainfall in Paraguay. But maybe all of *these* have been somehow rigged too! Of course, such a purely abstract proposal cannot be *refuted*, but besides being insane, it too would undercut scientific method. All scientific interpretations of our observations presuppose that they have not have been manipulated in such a way.

Finally, there is one big idea. Bell showed that measurements made far apart cannot regularly display correlations that violate his inequality if the world is local. But this requires that the measurements *have* results in order that there *be* the requisite correlations. What if no "measurement" ever *has* a unique result at all; what if *all* the "possible outcomes" occur? What would it even mean to say that in such a situation there is some *correlation* among the "outcomes of these measurements"?

This is, of course, the idea of the Many Worlds interpretation. It does not *refute* Bell's analysis, but rather *moots* it: in this picture, phenomena in the physical world do not, after all, display correlations between distant experiments that violate Bell's inequality, somehow it just *seems* that they do. Indeed, the world does not actually conform to the predictions of quantum theory at all (in particular, the prediction that these sorts of experiments have single unique outcomes, which correspond to eigenvalues), it just *seems* that way. So Bell's result cannot get a grip on this theory.

That does not prove that Many Worlds is local: it just shows that Bell's result does not prove that it isn't local. In order to even address the question of the locality of Many Worlds a tremendous amount of interpretive work has to be done. This is not the place to attempt such a task.

Finally, it has become fashionable to say that another way to avoid Bell's result and retain locality is to abandon *realism*. But such claims never manage to make clear at the same time just what "realism" is supposed to be and just how Bell's derivation presupposes it. I have heard an extremely distinguished physicist claim that Bell presupposes realism when he uses the symbol λ in his derivation. Here is how Bell characterizes the significance of λ [1] 15:

> Let this more complete specification be effected by means of parameters λ. It is a matter of indifference in the following whether λ denotes a single variable or a set, or even a set of functions, and whether the values are discrete or continuous.

There is obviously no physical content at all in the use of the symbol λ here. Bell makes no contentful physical supposition that can be denied.

There is no room for serial investigation of the various odd things that have been said about how Bell's proof has "realism" as a presupposition. The only way to proceed is first to reflect carefully on the arguments given above by Einstein and Schrödinger and Bell, and try to find any tacit physical claim that can be somehow denied. The Many Worlds interpretation provides an example, not merely of an assumption Bell tacitly makes but rather of an assumption most people make (viz. that experiments have unique outcomes) when considering the significance of Bell's result for the real world. Bell's theorem is only of relevance to theories that reproduce the predictions of standard quantum theory (or predict other violations of his inequality) for distant experiments. If "denying realism" means denying that no such predictions are actually correct, then of course Bell's theorem gives us no insight into such a view. But if one denies that the standard predictions are correct, there are more urgent interpretive matters to attend to than locality.

**Quo Vadimus?**

In an ideal world, a paper written on the 50[th] anniversary of a monumental theoretical result would be dedicated to reviewing how the result has shaped our picture of the world in the meantime, not what the result actually was. Unfortunately, we do not live in such a world, and the most urgent task even now is to make Bell's achievement clear. If we accept that the predictions of "standard quantum mechanics" are indeed accurate, then the world we live in is non-local. The first thing that has to be done is to accept that.

Acceptance is just the beginning. The next question should be: how is this non-locality implemented in a precisely defined physical theory? The problem of "standard quantum mechanics" not being a precisely defined theory, not up to "professional standards" for mathematical physics (which Bell also eloquently lamented), immediately takes center stage. Must a precisely defined theory that predicts violations of Bell's inequality for experiments at space-like separation postulate more space-time structure than is found in Relativity? That important question has been answered in the negative by construction, for example by the "Relativistic flashy GRW" theory of Roderich Tumulka. [21]. More urgently, if some approaches to incorporating non-locality into physics require modification of the Relativistic account of space-time, is that an unreasonable option to contemplate? My own view is that it is not, and that belief in the *completeness* of Relativity as an account of space-time structure has been irrationally fetishized just as belief in the *completeness* of the quantum-mechanical description had been by Bohr and company. With luck, every paper written on the 100[th] anniversary of Bell's theorem can safely focus on questions like these.

But not yet.

Acknowledgements

Many thanks for comments and feedback from David Albert, Jean Bricmont, Sean Carroll, Detlef Dürr, Shelly Goldstein, Ned Hall, Michael Kiessling, Dustin Lazarovici, Joel Lebowitz, Hannah Ochner, Howard Wiseman and Nino Zanghì.